\documentclass{sig-alternate}
\usepackage{graphicx}
\usepackage{subfig}
\usepackage[obeyspaces]{url}
\usepackage{balance}

\conf{Internetware '15}
\confinfo{November 6 2015, Wuhan, China}

\begin{document}

\title{A Study on Power Side Channels on Mobile Devices}
%\title{Covert Power Side Channels on Mobile Devices}

\numberofauthors{1}
\author{
\alignauthor
\large{Lin Yan, Yao Guo, Xiangqun Chen, Hong Mei} \\
\normalsize{Key Laboratory of High-Confidence Software Technologies (Ministry of Education), \\ School of Electronics Engineering and Computer Science, Peking University, Beijing, China} \\
 \normalsize{\{y.l, yaoguo,cherry, meih\}@pku.edu.cn}
}

\maketitle
\abstract{
Power side channel is a very important category of side channels, which can be exploited to steal confidential information from a computing system by analyzing its power consumption.
In this paper, we demonstrate the existence of various power side channels on popular mobile devices such as smartphones. Based on unprivileged power consumption traces, we present a list of real-world attacks that can be initiated to identify running apps, infer sensitive UIs, guess password lengths, and estimate geo-locations. These attack examples demonstrate that power consumption traces can be used as a practical side channel to gain various confidential information of mobile apps running on smartphones. Based on these power side channels, we discuss possible exploitations and present a general approach to exploit a power side channel on an Android smartphone, which demonstrates that power side channels pose imminent threats to the security and privacy of mobile users. We also discuss possible countermeasures to mitigate the threats of power side channels.
}




\section{Introduction}
%Side channels: timing, power analysis, etc.
Side channel attacks have drawn a lot of research attention for a long time. Originally, side channel attacks refer to attacks aimed to break a cryptosystem relying on physical information, which is neither the plain-text nor cipher-text. Recently, the context of side channel attacks has been expanded from cryptosystems to all kinds of computing systems. 
The goal of side channel attacks is gaining confidential information from the targeted computing system, while leveraging ``side channel'' information. Previously discovered side channels include timing information~\cite{Bortz:2007, Chen:2010}, sound~\cite{Zhuang:2009, Das:2014}, electro-magnetic radiation~\cite{Vuagnoux:2009}, shared memory/registers/files between processes~\cite{zhou:2013, Zhang:2009}, sensor information~\cite{placeraider-ndss13, schlegel:2011} and power consumption~\cite{kocher1999differential, MichalevskyNSB15}, etc.

\subsection{Power Side Channels}
Power analysis attacks (or \emph{power side channels}, PSCs) have become an important type of side channel attacks. The most famous example of power analysis attacks is the recovery of an encryption key from a cryptosystem~\cite{kocher1999differential, kocher:introduction}. Messerges \emph{et al.}\cite{messerges2002examining} examined both \emph{simple power analysis (SPA)} and \emph{differential power analysis (DPA)} attacks against the data encryption standard algorithm and managed to breach the security of smart-cards using signal-to-noise ratio (SNR) based multi-bit attack.

On mobile devices such as smartphones, Michalevsky \emph{et al.} proposed PowerSpy~\cite{MichalevskyNSB15}, which investigates the relationship between signal strength and the power pattern of smartphones. They showed that it is possible to infer smartphone users' whereabouts based on the power traces.

Our work also focuses on the mobile platform, but we will present a more comprehensive study of different PSCs on mobile devices.

\subsection{Attacks on Smartphones}

%Smartphone security, privacy issues...
Mobile devices including smartphones and tablets have become a popular computing platform. Mobile users are storing more and more privacy information such as passwords, credit card numbers, geo-locations, contacts information and even biometric information like fingerprints. Unfortunately, these sensitive data are vulnerable to various attacks, as evidenced by a huge number of malware aiming at stealing user privacy on Android devices~\cite{zhou2012dissecting}.

One popular example is UI-based attacks. For example, ScreenMilker~\cite{lin2014screenmilker}  can take screenshots of the foreground app covertly thus stealing user credentials.
Chen \emph{et al.} proposed an attack on the Android platform called UI inference attack~\cite{chen:2014}. They use the shared-memory side channel to infer UI states, in order to detect the correct timing for attacks.

There are many other attacks on mobile devices, we will present several of them that can be enabled by exploiting \emph{only} power traces of a device.

%\vspace{-0.05in}
\subsection{Summary and Contributions}

In this paper, we present our study on the existence of \textbf{\emph{power side channels}} (PSCs) on mobile devices, which may bring serious threats to mobile user security and privacy. We also demonstrate some potential attacks based on these PSCs.

%Experiments:
%- measurements by power monitor.
%- experiments with real power traces collected on smartphones.
We conduct our study on PSCs through a series of experiments to collect power patterns of mobile devices. The power patterns can be collected through both hardware-based method (with a Monsoon Power Monitor~\cite{monsoonPM} attached to the smartphone) and software-based method (directly polling voltage and current readings within the mobile system). Then we analyze collected power traces to see if the uniqueness of different characteristics of the confidential data can be reflected on power patterns.
%Results from both methods demonstrate the repeatability of the power pattern from the same UI and the distinguishability between power patterns from different UIs within a mobile app, thus revealing the existence of power side channels.

With experiments on a Nexus 5 smartphone, we demonstrate several prominent PSCs on mobile devices that can be exploited to gain various kinds of confidential information. Besides previously reported PSCs used to estimate the geo-location of mobile devices~\cite{MichalevskyNSB15}, we also identify the following new PSCs that can be used to: (1) identify different mobile apps (based on power patterns during app loading); (2) infer different UIs within one mobile app (based on collected sensitive UI traces); and (3) guess password lengths (based on power patterns during password input).

Based on these newly discovered PSCs, we discuss possible exploitations and present a comprehensive exploitation scenario, in which the malicious party could steal user passwords by leveraging PSCs in a particular order. To show how the exploitation works in reality, we present an overview of a general approach to exploit PSCs on Android smartphones. 
%We present a case study based on the approach and the results show that the PSC can be successfully exploited to infer sensitive UIs and help steal passwords in mobile apps. 

Our contributions can be summarized as follows.
\begin{itemize}
%\vspace{-0.08in}
\item We present the first demonstrative study to reveal the existence of multiple power side channels on smartphones.
%\vspace{-0.08in}

\item We demonstrate that power side channels pose practical threats as they can be used to fulfill different attack purposes, such as app identification, UI identification or password guessing.

\item We present a general approach to exploit power side channels.
%\vspace{-0.08in}

\item We discuss possible countermeasure techniques to mitigate the threats of power side channels.
    %such as making the reading of power patterns privileged on mobile devices, or at least revealing more coarse-grained data to apps (only battery levels, no detailed power numbers, etc.), or obfuscating the power pattern.

 \end{itemize}

The rest of this paper is organized as follows. We first introduce our methodology in Section \ref{sec:method}. In section \ref{sec:pscs}, we present various power side channels we discover on mobile devices. Section \ref{sec:exploitation} presents the details of our research on the exploitation of  power side channels. We discuss our limitations and possible defenses and future work in Section \ref{sec:dis}. Background of our work and survey of related work are presented in Section \ref{sec:relate} and Section \ref{sec:con} concludes this paper.

%\vspace{-0.05in}

\section{Methodology}
\label{sec:method}

In this section, we introduce the experimental setup and how we collect power traces from a smartphone.

%\vspace{-0.05in}
\subsection{Experimental Setup}

We use a Google Nexus 5 smartphone to perform most of the experiments. The power numbers can be measured with a Monsoon Power Monitor with a stable 4.2V voltage. We install popular apps on Android ASOP 4.4 on the testing phone and measure the power numbers in different cases.

%\vspace{-0.05in}
\subsection{Hardware Measurement}

Hardware measurement is straightforward. We use the Monsoon Power Monitor~\cite{monsoonPM} in our study to supply a stable voltage to smartphones and measure the current value in various stages. The power monitor works as an accurate hardware power meter, which offers stabilized power supply to the smartphone being tested and records the power consumption numbers of the smartphone in real-time. Since the voltage is a constant, we simply use current value to denote the power consumption.

%Due to the huge amount of data generated by the Power Monitor, we are unable to perform analysis on a long trace such as several hours. Instead, based on our analysis on traces, we selectively monitor the power current on various representative short periods.

%With the current curves, we could find some power patterns of the standby mode. Together with the event traces, we could tell the reason of these patterns.

%\vspace{-0.05in}

\subsection{Software Measurement}
\label{sec:softmethod}

Although power measurements based on hardware meters can be used to confirm the existence of PSCs, it might still not be applicable to real-world exploitation since the hardware meter is apparently not a standard auxiliary accessory for smartphones.

Fortunately, power related statistics are publicly accessible on most smartphone OSes such as Android. In general, instant power numbers can be calculated based on voltage and current readings of BMU (Battery Monitoring Unit)~\cite{yoon2012appscope}. The battery status information is accessible by most apps without system-level privilege. %On many occasions, mobile apps need to know the battery status to carry out responses such as saving user context before the battery dies.

In our experiments, we develop a polling collector on a Nexus 5 smartphone which reads and records power patterns by polling battery status files, which are listed in Table \ref{tab:files}.

\begin{table}
\centering
\caption{Battery status files used in our experiments (on the Android platform for Nexus 5).}
\label{tab:files}
\begin{tabular}{| p{1.4in} | p{1in} | c |}
\hline
File location & Description & Unit\\
\hline
\path{/sys/class/power_supply/battery/voltage_now} & Instant voltage reading of battery &  \ensuremath{\mu V}\\
\hline
\path{/sys/class/power_supply/battery/current_now} & Instant current reading of battery &  \ensuremath{\mu A}\\
\hline
\end{tabular}
%\vspace{-0.15in}

\end{table}

%\vspace{-0.05in}
\subsection{Identification of Power Side Channels}

In this paper, a \emph{power side channel} (PSC) is defined as an identifiable power pattern that can be used to infer the secret information from a running app on a mobile device.

We assume that power consumption traces are available as an unprivileged source that can be obtained by any normal apps running on a mobile device, which is the case for most Android devices.

The purpose of this paper is to demonstrate the existence of PSCs on mobile devices that can be exploited to perform various attacks. We achieve this by providing experimental results that show repeatable and unique power traces that can be detected during runtime. We also cover the details on how to exploit these PSCs to conduct real attacks. We believe that these PSCs pose real threats to the security and privacy of mobile users.

%Existence of side channels on smartphones. examples: location, UI states (app loading and sensitive UIs).
%This paper: study on the existence of power side channels and potential exploitation techniques.

%\vspace{-0.05in}
\section{Power Side Channels}
\label{sec:pscs}

This section presents our findings on potential power side channels (PSCs) on mobile devices, especially three new PSCs that can be exploited to identify running apps, infer sensitive UIs and guess password information, respectively.

%\vspace{-0.05in}
\subsection{PSC \#1: App Identification}
\label{sec:appid}

For many real-world attacks aiming at mobile systems, knowing which app is running in the foreground is important. Knowing the identity of the foreground app means being aware of the timing information of the attacking target. Previously, identity of the foreground app can be accessed by calling the \texttt{getRunningTasks()} method of the \texttt{ActivityManager} on Android for all apps. However, this method is officially unavailable to third-party apps after Android 5.0 due to security concerns~\cite{androidAM}.

\begin{figure*}[!t]
\centering
\begin{tabular}{ccc}
\subfloat[]{\includegraphics[width=1.4in]{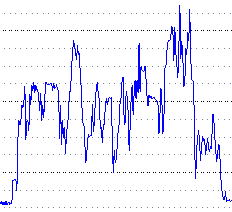}} &
\subfloat[]{\includegraphics[width=1.3in]{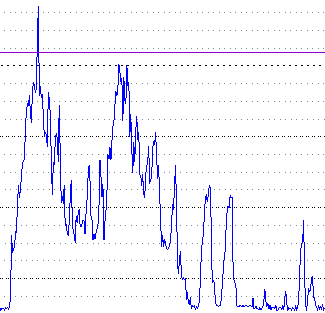}} &
\subfloat[]{\includegraphics[width=0.8in]{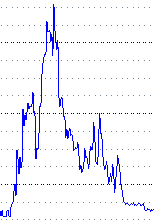}}\\
\subfloat[]{\includegraphics[width=1.4in]{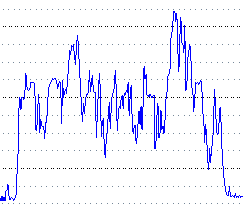}} &
\subfloat[]{\includegraphics[width=1.3in]{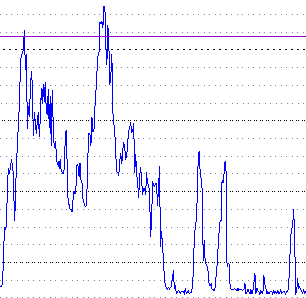}} &
\subfloat[]{\includegraphics[width=0.8in]{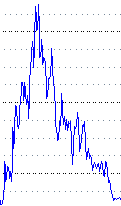}}\\
\end{tabular}
\caption{Hardware-based measurements on the loading phases of different apps in two test runs: (a), (d) for WeChat; (b), (e) for Alipay; (c), (f) for Gmail.}
\label{fig:app_load}
%\vspace{-0.15in}
\end{figure*}

Based on our findings, the foreground app can also be identified by analyzing the power patterns of the app loading phase. We choose three popular apps from three common app categories(social media, e-commerce and productivity) and record the power patterns of the loading phases. The results are shown in Figure \ref{fig:app_load}. It is obvious that different apps exhibit distinguishable power patterns and different test runs of the same app generate visually similar power curves.  

We have conducted more than two dozen different test runs on these apps to test the repeatability of the power patterns during app loading. The results show similar patterns as the data shown in Figure 1. However, due to space limit, we cannot show all the results in this paper.

On the other hand, we have also tested the app to collect its power pattern when the apps continue to run after the app loading phase. The results show that the power patterns during the app loading phase is distinguishable from power traces in other execution phases.

We also collected power traces during app loading using the software-based method and got similar results.
%The results are similar and omitted due to space limitation. 
Because software-based power traces are more practical for real-world attacks, we will present only software-based traces in the next case.

%\vspace{-0.05in}
\subsection{PSC \#2: UI Inference}
\label{sec:uiin}

The UI states within a mobile app are also crucial for many attacks. For example, in order to initiate activity hijacking attacks on Android, attackers need to know when the user login UI will be prompted so that they can intercept the UI state transitions and insert fake user login UIs that could steal user credentials.

However, what is going on inside an app is not directly observable by other parties without system-level privileges if the app chooses not to share it. A previous study has shown that attackers could peek into mobile applications via the shared-memory side channel along with some UI signatures based on CPU utilization time and other runtime events~\cite{chen:2014}.

\begin{figure*}[!t]
\centering
\includegraphics[width=5.8in]{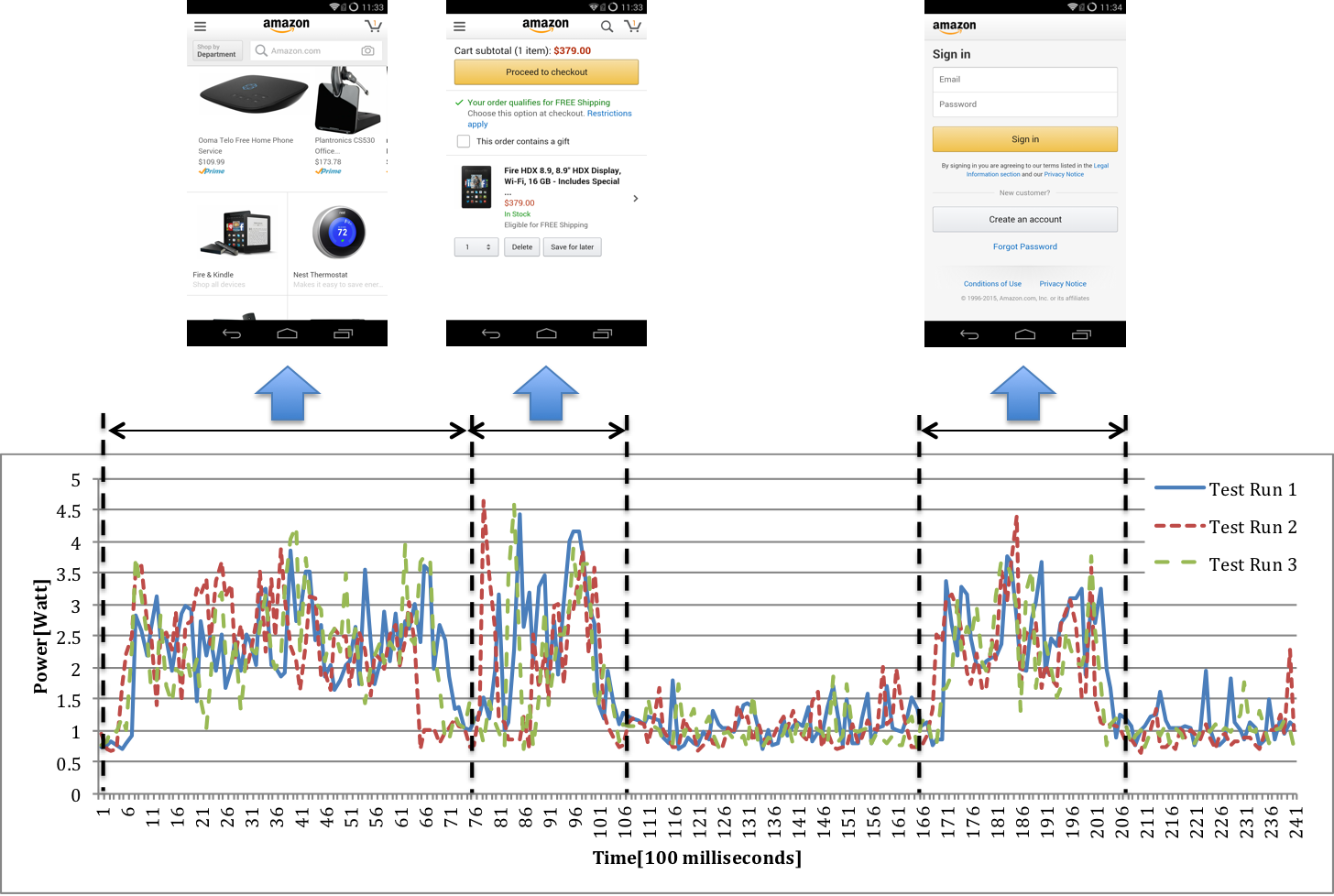}
\caption{Power traces collected through software-based measurement for the Amazon app. The three UIs shown above are \emph{app entry, product details, and user login}, respectively.}
\label{fig:amazon_software}
%\vspace{-0.15in}
\end{figure*}

Different UI states within one mobile app can also be inferred through their respective power traces. As depicted in Figure \ref{fig:amazon_software}, we conduct experiments on collecting power patterns of visiting three different UIs in the Amazon app in multiple test runs. The power patterns demonstrate both the repeatability of a single UI and the distinguishability between different UIs within the same mobile app.

For example, in order to infer the login screen of a mobile app, an attacker can first collect the power patterns of the targeted UI in advance on another device. After learning the power pattern, it is possible to detect the pattern on the targeted device with a pattern matching algorithm. In this way, the attacker would be able to infer when the targeted UI is loaded on the targeted device, i.e., recognizing the exact timing to conduct attack.

%\vspace{-0.05in}
\subsection{PSC \#3: Password Length Guessing}

For a more complex scenario, we attempt to explore whether power traces can be used to reveal information on user passwords, similar to previous power analysis approaches on other platforms.

\begin{figure*}[t]
\centering
\includegraphics[width=5.6in]{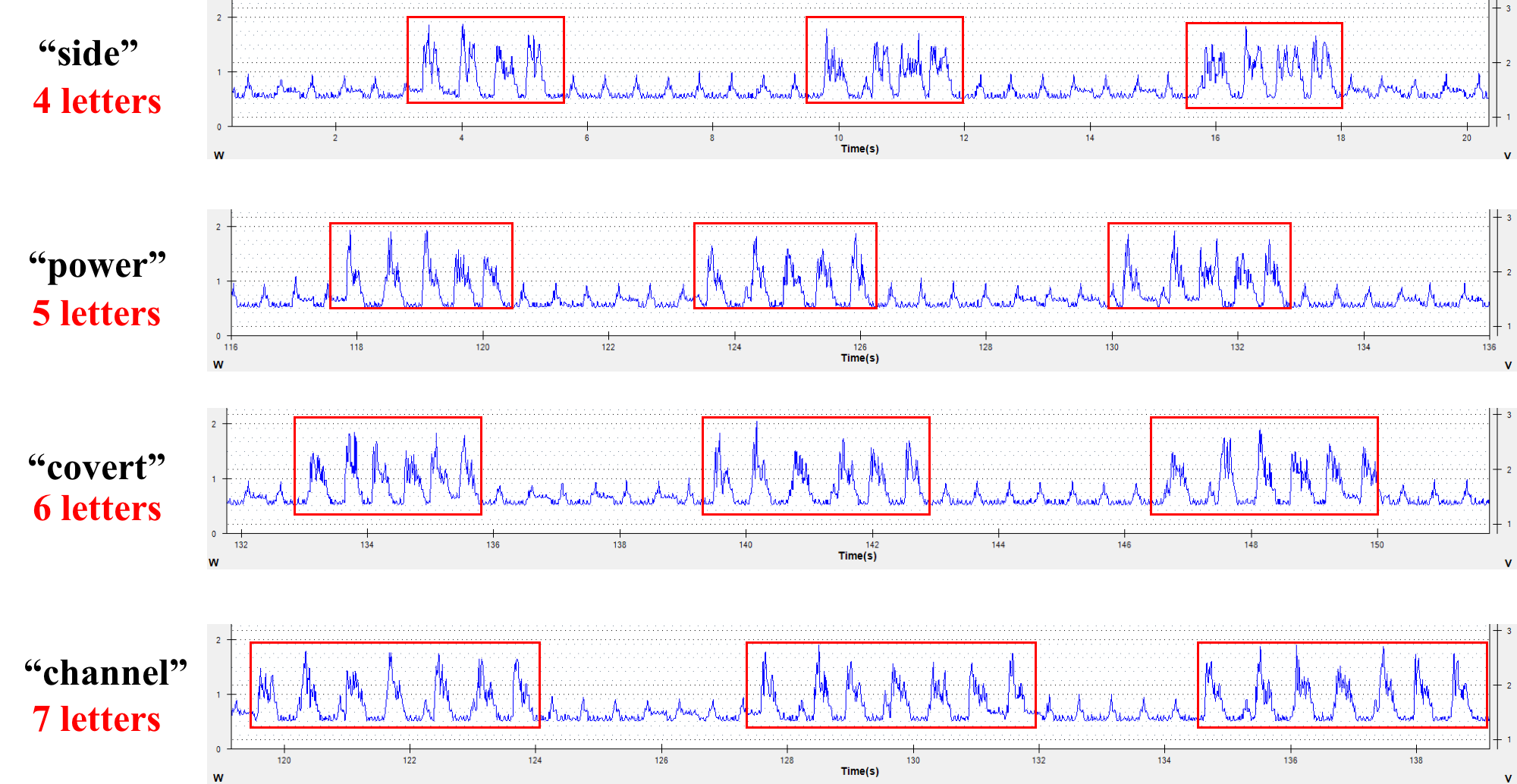}
\caption{Power patterns from hardware measurements when inputting passwords with different lengths.}
\label{fig:wordlength_pm}
%\vspace{-0.15in}
\end{figure*}

Although it is difficult to guess the actual passwords, we find that it is relatively easy to identify the length of a password during password input. Figure \ref{fig:wordlength_pm} depicts the hardware-measured power traces during password input. Each time the user inputs a single letter of the passwords, it will be reflected in the power pattern as a sudden ``burst'' of power consumption. It is obvious that the length of user's password can be inferred by counting the number of consecutive power pattern ``bursts''.

Obtaining the length of a user password is meaningful for attackers to guess the password since it reduces the complexity of the cracking algorithm, as well as the size of the helping dictionary.

%\vspace{-0.05in}
\subsection{PSC \#4: Geo-location Estimation}

The geo-location of a mobile device can also be revealed by its power consumption pattern. A recent study uses machine learning techniques to identify the routes taken by the smartphone users based on previously collected power consumption data~\cite{MichalevskyNSB15}.

There are many directions can be followed on this thread of work. For example, based on our observation, the power consumption patterns can be used to track the accurate real-time location of a mobile device, if a database of power patterns on location signatures has been collected in advance.

%\vspace{-0.05in}

\section{PSC Exploitations}
\label{sec:exploitation}

In this section we present a detailed description of how to exploit power side channels (PSCs) on mobile devices. We first discuss possible exploitations on mobile devices leveraging presented PSCs. Then we introduce a general PSC Exploitation approach. 
%Finally we present a case study of exploiting the PSC to infer sensitive UIs.

\subsection{Possible PSC Exploitations}
PSCs on mobile devices can be exploited by malicious parties to steal user privacy information such as login credentials and geo-locations. 

\begin{figure}[!t]
\centering
\includegraphics[width=3.2in]{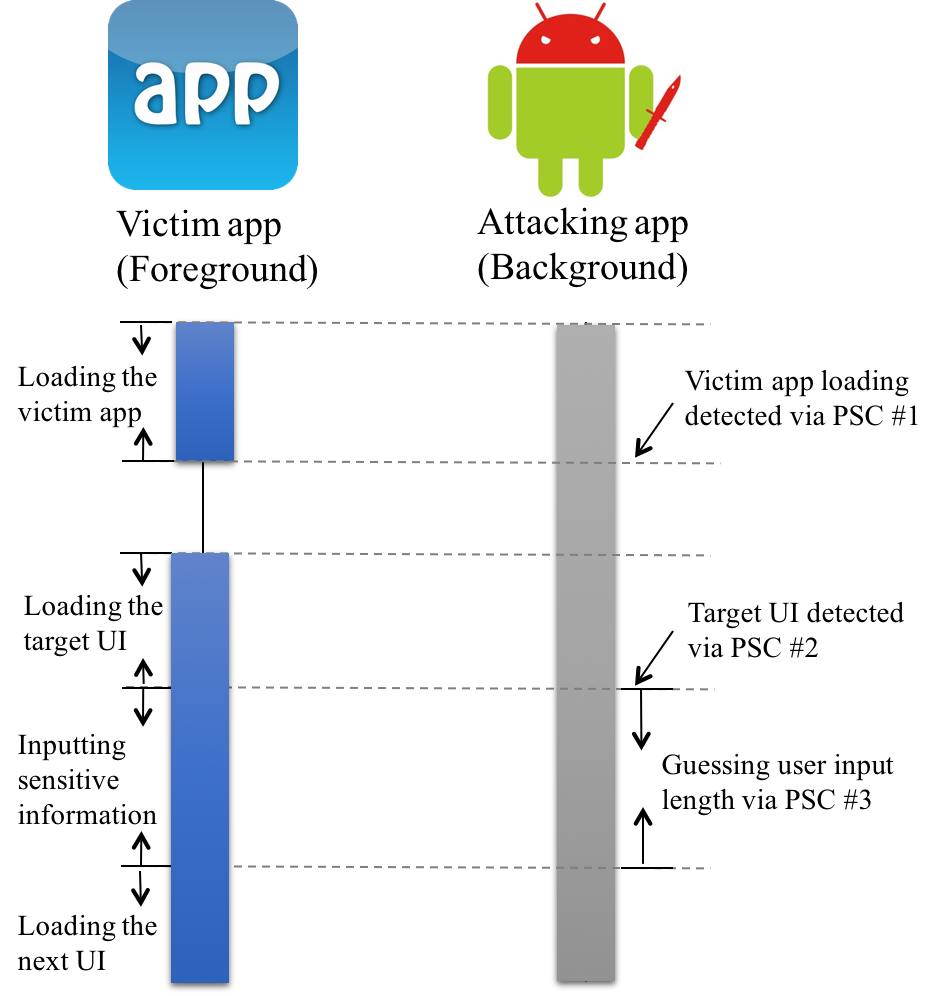}
\caption{Possible Exploitations of presented PSCs.}
\label{fig:attack}
%\vspace{-0.15in}
\end{figure}

For example, as depicted in Figure \ref{fig:attack}, PSCs \#1, \#2 and \#3 found in the previous section can be combined to guess user login credentials of a certain app. To be more specific, the attack can first learn the power patterns of the victim app, the target UI and the user input. Then the attacker can develop a malicious app to fulfill PSC exploitations.  The malicious app can first detect the loading of the victim app via PSC \#1; then the malicious app can capture the occurrence of the login UI via PSC \#2; finally the malicious app can help crack user password with the password length information acquired via PSC \#3. It should be noted that the malicious app does not need any special permission to read system power patterns on popular mobile systems like Android.

The key of the malicious apps to exploit PSCs is to achieve automated detection of some pre-learned power patterns. This can be done by adopting pattern matching or machine learning algorithms, such as dynamic time warping (DTW) \cite{muller:dynamic}, which can be used to match a previously learned pattern in a continuous power trace. However, the pattern matching process may affect the total system power consumption during detection, thus it should be implemented as a light-weight detection process if used in real-time.

In the above exploitation scenario, all attacks are based on PSCs. However, it is not necessary to use the power channel alone to conduct the attack. Attackers can choose to combine PSCs with other attacking channels to improve the attacking success rates. 
On particular example is that an attacker can choose to take screenshots continuously after detecting the target UI and apply image recognition algorithms to retrieve user passwords in the screenshots. Apparently,  screenshot-based attacks are orthogonal to PSC based attacks we discussed and out of scope of this paper.  

\subsection{A General Exploitation Approach}
\begin{figure*}[t]
\centering
\includegraphics[width=5.6in]{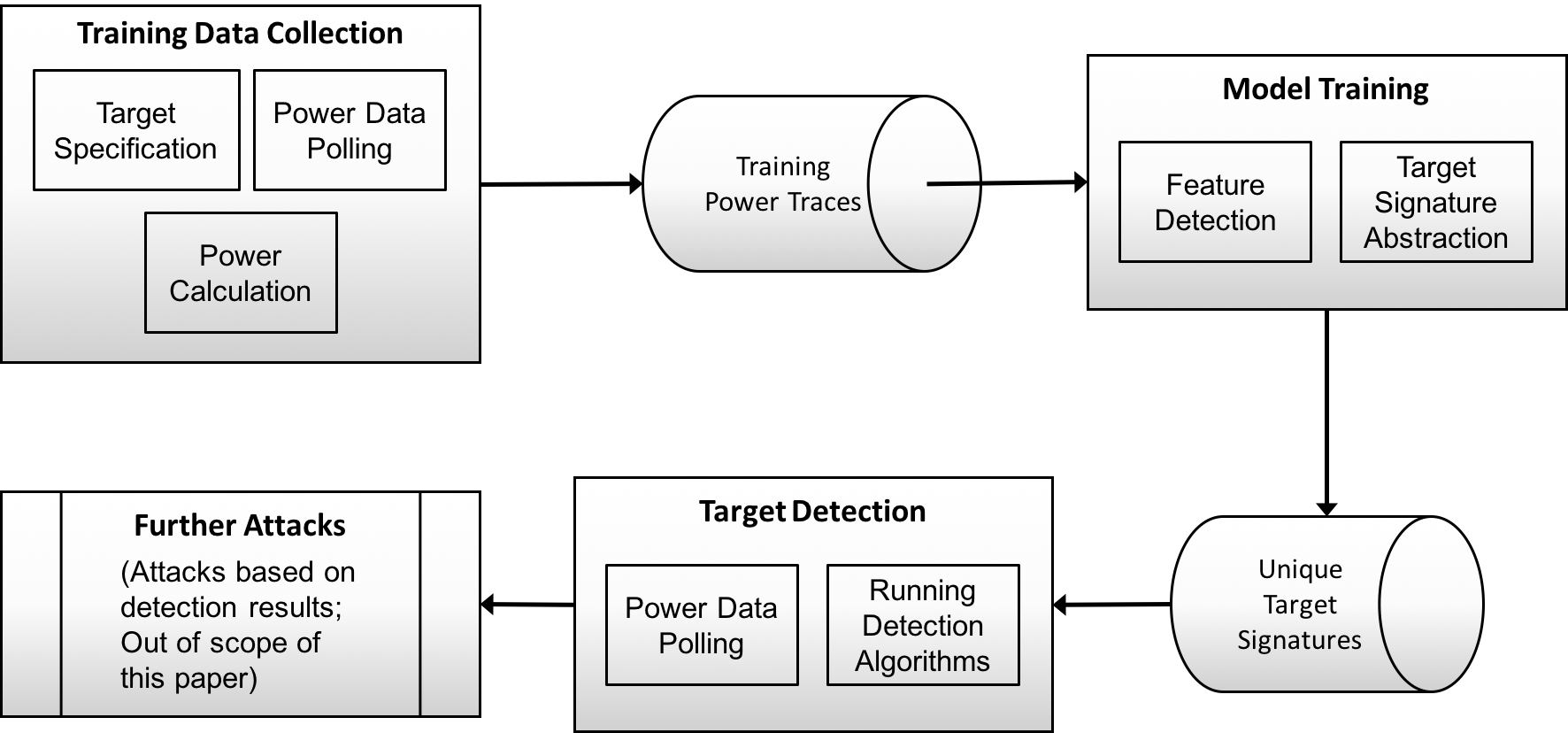}
\caption{An overview of a general PSC exploitation approach.}
\label{fig:overview}
\end{figure*}

Figure \ref{fig:overview} shows the overview of a general PSC exploitation approach. The purpose of PSC exploitation is to find an effective method to detect the occurrences of the target power pattern from a continuous power trace collected from an app.
PSC exploitation involves the following main steps.

%\vspace{-0.1in}
\subsubsection{Training Data Collection} 

The first step in collecting training data is selecting a target. The choice of the target depends on which kind of PSC the attacker wants to exploit. For example, the loading phase of a mobile app might be the target if the attacker wants to leverage PSC \# 1 described in Section \ref{sec:appid}. Or if the attacker wants to leverage PSC \# 2 as shown in Section \ref{sec:uiin}, the loading of sensitive UIs are considered to be the target.

After choosing the target, the attacker must specify the start and end of the target explicitly so that its power pattern can be precisely measured. As we have introduced in Section \ref{sec:softmethod}, the voltage and current numbers of running the target can be acquired by polling unprivileged system power files and the power can be calculated based on them.

%\vspace{-0.05in}
\subsubsection{Model Training} Based on the collected power traces of the target, the attacker needs to generate a signature to represent the target. This step generally involves detecting features of the collected power traces and and signatures accordingly. The specific methods for detecting features abstracting the signature depend on the characteristics of the training dataset and PSC exploitation scenario. 

%\vspace{-0.05in}
\subsubsection{Target Detection} We assume that an attacking app runs in the background on the smartphone, which continuously collect the power traces. Based on the signature of the target trained in previous steps, the attacking app is able to detect the occurrence of the target in real-time by running appropriate detection algorithms. Detection algorithms are pattern matching algorithms in essence and the choice of them depends on characteristics of the PSC exploitation scenario.

\section{Discussions}
\label{sec:dis}

This section discusses the limitations and countermeasures on the PSCs identified in the previous section.
\subsection{Limitations}
Although we have demonstrated the existence of multiple PSCs. There are some limitations in the current study, most of which we plan to explore in future work:
%\vspace{-0.25in}

\begin{itemize}

\item Identifying the side channel is only the first step. In order to perform real-world attacks using these side channels, there are various issues that should be considered, such as how to collect power patterns accurately, how to detect app loading or sensitive UIs in real-time, how to apply successful attacks in practical scenarios, etc.
%\vspace{-0.05in}

\item One particular issue that needs to investigate is how to guarantee that the power patterns keep unchanged during real-time detection. For example, a pattern matching app should be carefully implemented such that it does not cause significant power overheads during real-time detection. Otherwise it will difficult to identify the power patterns due to power overhead of the detecting process itself.
%\vspace{-0.05in}

\item In our experiments, we have only tested the ideal case in which the background power consumption noise is minimal. In reality, there might exist multiple apps running in the background. Because these background apps might make network or I/O requests in an irregular way, it may affect the power patterns when we tried to detect the patterns revealed by the PSCs. This issue may also increase the difficulty to exploit discovered PSCs, nonetheless PSCs are practical threats on mobile devices.

\end{itemize}

\subsection{Possible Mitigating Techniques}
\begin{itemize}
%\vspace{-0.15in}
\item
\textbf{Energy obfuscation through code injection.} One straightforward mitigation approach is that we can inject meaningless code into mobile apps, in order to insert power bursts into its power pattern to make it unpredictable. This can be achieved at the source-code level during app development, or through instrumentation to the bytecode for app binaries.
%\vspace{-0.05in}

\item
\textbf{Randomly changing display/color parameters.}  One interesting feature for the OLED or AMOLED displays used for smartphones is that it consumes different power when different color schemes are used~\cite{dong2012oled}. Thus we can vary the displaying color and other parameters during the execution of apps we want to protect. This could also be achieved during app development or through bytecode instrumentation~\cite{Li:2014:MWA}.

%\vspace{-0.05in}
\item
\textbf{Raising the privilege needed to access power files.} Of course, we can always make the power information privileged, such that not all apps could access these data directly. As a matter of fact, mobile apps probably do not need to read low-level power related files containing raw voltage or current readings. The only thing that most apps need to know is how much battery is still remaining, which should not pose serious threats as a side channel.
\end{itemize}

\section{Background and Related Work}
\label{sec:relate}

%\textcolor[rgb]{1,0,0}{ADD More details, and more references.}

In this section, we introduce background knowledge on power characteristics of Android smartphones and related research work on side channel attacks, especially power side channel attacks.
%\vspace{-0.1in}
\subsection{Power Characteristics of Android Smartphones}

In this paper, we use Android smartphones as our studying subject as Android is currently the most popular mobile operating system for smartphones and tablets. Similar algorithms could be applied on Android tablets, as they share the same framework and most of the APIs.

For a different mobile OS such as Apple iOS, the power consumption patterns of their mobile apps share very similar characteristics as Android because they use similar hardware components and similar app structure. Thus the attacks could potentially be repeated on iPhones as well.

Most smartphones are power-hungry devices, while many batteries could last only less than a day in typical usage. The most power consuming components are CPU, network, screen and various sensors such as GPS and camera~\cite{carroll2010analysis}. For most Android smartphones, the screen consumes a majority of the total power. The screen power could be affected significantly due to its brightness and color of pixels~\cite{dong2012oled}, thus it is able to reduce the power consumption of an app by adjusting its color schemes~\cite{li2014making}.

For a mobile application, the power of a particular execution trace in a certain stable environment is determined by the behavior of the app. Different types of apps consume different power according to their usage of CPU cycles, network traffic and screen brightness, etc. Power consumption for an app can therefore be modeled by their resource usages~\cite{yoon2012appscope}, system call traces~\cite{yetim2012eprof} and source code ~\cite{hao2013estimating}, etc.

Besides power consumed when an app runs in the foreground (i.e., with user interaction), many apps also consume significant power when their respective services run in the background, for example, checking new emails or new tweets regularly. Although background power could potential drain a lot of battery in the long run, it is relative small compared to the power consumed by foreground apps. In this paper, we only consider the power patterns of an app when it runs in the foreground.

\subsection{Side Channel Attacks}
Side channel attacks have drawn researchers' attention for decades.  Wright and Greengrass \cite{wright:spycatcher} reported the first official record on utilizing side channel attacks that happened in 1965.  At first, research on side channel attacks mainly focused on cryptography \cite{kocher:timing, biham:differential, kocher:introduction} and encrypted communications~\cite{Song:2001, Brumley:2003, Saponas:2007, Wright:2008}. In these study, side channel attacks are defined as successfully decrypting the ciphertext based on the information gathered from the encryption system that is neither the plain-text being encrypted nor the cipher-text after encryption.

Along with the research development, side channel attacks are discussed more broadly in many contexts, not necessarily limited to the contexts of cryptography and encrypted communications. Previously discovered common side channels include timing channels~\cite{wray1991analysis, gianvecchio2007detecting, cabuk2004ip, Bortz:2007, Chen:2010, Kopf:2010, Zhang:2012}, acoustic channels~\cite{Zhuang:2009, Das:2014}, electromagnetic waves~\cite{agrawal2003side, Vuagnoux:2009}, shared memory/registers/files between processes~\cite{Qian:2012, jana:2012, zhou:2013, Zhang:2009}, sensor metrics~\cite{Xu:2012, Miluzzo:2012, placeraider-ndss13, schlegel:2011} and power consumption~\cite{kocher1999differential, MichalevskyNSB15}. Types of information can be leaked via these side channels include almost anything related to sensitive information about users, such as keystrokes~\cite{Killourhy:2009}, locations~\cite{MichalevskyNSB15}, speech~\cite{michalevsky2014gyrophone} or even health data~\cite{Chen:2010}.

In order to protect computing systems from side-channel attacks, many research work have proposed various mitigation or protection methods. Examples include redesigning the encryption methods to prevent power analysis~\cite{moller2001securing, brier2002weierstrass}, predictive mitigation of timing channels for interactive applications~\cite{zhang2011predictive}, system-level protection against cache-based side channel attacks~\cite{kim2012stealthmem}, and new cache designs to thwart software cache-based side channel attacks~\cite{wang2007new}.
%We focus on exploring the power consumption based side channel on mobile platforms in our work. %The root cause of this side channel is because power patterns is publicly observable to all mobile apps.%

\subsection{Power Side Channels}
Power analysis attacks (or power side channels)~\cite{brier2004correlation} have become an important type of side channel attacks in recent years. One well-known example of power analysis is the recovery of an encryption key from a cryptosystem~\cite{kocher1999differential, kocher:introduction}. Messerges \emph{et al.}~\cite{messerges2002examining, messerges1999power} examined both simple power analysis(SPA) and differential power analysis (DPA) attacks against the data encryption standard (DES) algorithm and managed to breach the security of smart-cards using the proposed signal-to-noise ratio (SNR) based multi-bit attack.\cite{mangard2008power}

In order to defeat power analysis attacks on smart-cards, Herbst \emph{et al.}~\cite{ herbst2006aes} present an efficient AES software implementation that is suited for 8-bit smart-cards and resistant against power analysis attacks. Ratanpal \emph{et al.}~\cite{ratanpal2004chip} presents a circuit that can be added to crypto-hardware to suppress information leakage through the power supply pin side channel.

Chari \emph{et al.}~\cite{chari1999towards} propose a sound approach to counteract power analysis attacks. It includes an abstract model which approximates power consumption in most devices and a generic technique to create provably resistant implementations for devices where the power model has reasonable properties. They prove a lower bound on the number of experiments required to mount statistical attacks on devices whose physical characteristics satisfy reasonable properties.

Power side channels have also been discovered on other systems besides smart-cards. For example, Hlavacs \emph{et al.}~\cite{hlavacs2011energy} demonstrate that energy consumption side-channel attack can be performed between virtual machines in a cloud.

On mobile platforms, Michalevsky \emph{et al.} proposed PowerSpy~\cite{MichalevskyNSB15}, which investigates the relation between signal strength and the power pattern of the smartphone and showed that they can infer smartphone users' whereabouts based on the power traces.

Our work also focuses on the mobile platform, but we have presented different and more general exploitations based on power traces.

%\subsection{Smartphone attacks}
%The UI security of an application has been studied extensively~\cite{Shapiro:2004, Feske:2005, chen2007systematic, Fischer2009}. On traditional desktop platforms, UI-based attacks are sometimes categorized as UI spoofing attacks~\cite{chen2007systematic, Fischer2009}.  
%
%Recently, UI-based attacks start to emerge on mobile platforms. For example, Capturer~\cite{xu2009stealthy} allows the attacker to automatically activate the built-in camera on Windows smartphones and take videos secretly, which will be used to conduct situation detection. ScreenMilker~\cite{lin2014screenmilker}  can take screenshots of the foreground app covertly and steal user credentials. SUPOR~\cite{huang2015supor} presents a static mobile app analysis tool for detecting sensitive UIs and associated variables in the app code.
%
%There are also UI-based attacks on smartphones leveraging physical side channel information. For example,  iSpy~\cite{raguram2011ispy} records a video of the victim's smartphone screen reflected on nearby objects such as the lens of victim's sunglasses, and then automatically reconstructs the texts typed on the virtual keyboard using machine learning techniques.
%
%Chen \emph{et al.} propose an attack on the Android platform called UI inference attack~\cite{chen:2014}. They use the share-memory side channel to infer UI states, in order to detect the correct timing for attacks.
%Our work targets at a similar attack in UI inference, but we have achieved it through power side channel exploitation.

\section{Concluding Remarks}
\label{sec:con}

In this paper, we have demonstrated the existence of various power side channels that can be exploited to perform attacks to steal private information about apps running on mobile devices. 

We first provide our discovery of several power side channels on mobile devices. Then we provide a general exploitation approach of these power side channels. %We show the feasibility of exploiting power side channels on mobile devices through a case study on smartphones. 

We believe these power side channels pose real threats to mobile user security and privacy if they are exploited in real-world attacks, thus needed to be addressed in an urgent manner.

%\vspace{-0.05in}
\section{Acknowledgement}
This work is supported by the High-Tech Research and Development Program of China under Grant No. 2013AA01A605 and  the National Natural Science Foundation of China under Grant No. 61421091 and No. 61103026.

\bibliographystyle{abbrv}
\balance
\bibliography{PSCMobile}

\begin{thebibliography}{10}

\bibitem{androidAM}
Android developers - {ActivityManager refernces}.
\newblock
  \url{http://developer.android.com/reference/android/app/ActivityManager.html#getRunningTasks(int)}.

\bibitem{monsoonPM}
{Monsoon Power Monitor}.
\newblock \url{https://www.msoon.com/LabEquipment/PowerMonitor/}.

\bibitem{agrawal2003side}
D.~Agrawal, B.~Archambeault, J.~R. Rao, and P.~Rohatgi.
\newblock The {EM} side channel (s).
\newblock In {\em Cryptographic Hardware and Embedded Systems-CHES 2002}, pages
  29--45. Springer, 2003.

\bibitem{biham:differential}
E.~Biham and A.~Shamir.
\newblock Differential fault analysis of secret key cryptosystems.
\newblock In {\em Advances in Cryptology (CRYPTO '97)}, pages 513--525.
  Springer, 1997.

\bibitem{Bortz:2007}
A.~Bortz and D.~Boneh.
\newblock Exposing private information by timing web applications.
\newblock In {\em Proceedings of the 16th International Conference on World
  Wide Web}, pages 621--628. ACM, 2007.

\bibitem{brier2004correlation}
E.~Brier, C.~Clavier, and F.~Olivier.
\newblock Correlation power analysis with a leakage model.
\newblock In {\em Cryptographic Hardware and Embedded Systems-CHES 2004}, pages
  16--29. Springer, 2004.

\bibitem{brier2002weierstrass}
E.~Brier and M.~Joye.
\newblock Weierstra{\ss} elliptic curves and side-channel attacks.
\newblock In {\em Public Key Cryptography}, pages 335--345. Springer Berlin
  Heidelberg, 2002.

\bibitem{Brumley:2003}
D.~Brumley and D.~Boneh.
\newblock Remote timing attacks are practical.
\newblock In {\em Proceedings of the 12th Conference on USENIX Security
  Symposium - Volume 12}. USENIX Association, 2003.

\bibitem{cabuk2004ip}
S.~Cabuk, C.~E. Brodley, and C.~Shields.
\newblock Ip covert timing channels: design and detection.
\newblock In {\em Proceedings of the 11th ACM conference on Computer and
  communications security}, pages 178--187. ACM, 2004.

\bibitem{carroll2010analysis}
A.~Carroll and G.~Heiser.
\newblock An analysis of power consumption in a smartphone.
\newblock In {\em USENIX annual technical conference}, volume~14, 2010.

\bibitem{chari1999towards}
S.~Chari, C.~S. Jutla, J.~R. Rao, and P.~Rohatgi.
\newblock Towards sound approaches to counteract power-analysis attacks.
\newblock In {\em Advances in Cryptology (CRYPTO 99)}, pages 398--412.
  Springer, 1999.

\bibitem{chen:2014}
Q.~A. Chen, Z.~Qian, and Z.~M. Mao.
\newblock Peeking into your app without actually seeing it: {UI} state
  inference and novel {Android} attacks.
\newblock In {\em Proceedings of the 23rd USENIX Conference on Security
  Symposium}, pages 1037--1052, Berkeley, CA, USA, 2014. USENIX Association.

\bibitem{Chen:2010}
S.~Chen, R.~Wang, X.~Wang, and K.~Zhang.
\newblock Side-channel leaks in web applications: A reality today, a challenge
  tomorrow.
\newblock In {\em Proceedings of the 2010 IEEE Symposium on Security and
  Privacy}, pages 191--206. IEEE Computer Society, 2010.

\bibitem{Das:2014}
A.~Das, N.~Borisov, and M.~Caesar.
\newblock Do you hear what {I} hear?: Fingerprinting smart devices through
  embedded acoustic components.
\newblock In {\em Proceedings of the 2014 ACM SIGSAC Conference on Computer and
  Communications Security}, CCS '14, pages 441--452. ACM, 2014.

\bibitem{dong2012oled}
M.~Dong and L.~Zhong.
\newblock Power modeling and optimization for {OLED} displays.
\newblock {\em IEEE Transactions on Mobile Computing}, 11(9):1587--1599, Sept
  2012.

\bibitem{gianvecchio2007detecting}
S.~Gianvecchio and H.~Wang.
\newblock Detecting covert timing channels: an entropy-based approach.
\newblock In {\em Proceedings of the 14th ACM conference on Computer and
  communications security}, pages 307--316. ACM, 2007.

\bibitem{hao2013estimating}
S.~Hao, D.~Li, W.~G. Halfond, and R.~Govindan.
\newblock Estimating mobile application energy consumption using program
  analysis.
\newblock In {\em 35th International Conference on Software Engineering
  (ICSE)}, pages 92--101. IEEE, 2013.

\bibitem{herbst2006aes}
C.~Herbst, E.~Oswald, and S.~Mangard.
\newblock An {AES} smart card implementation resistant to power analysis
  attacks.
\newblock In {\em Applied cryptography and Network security}, pages 239--252.
  Springer, 2006.

\bibitem{hlavacs2011energy}
H.~Hlavacs, T.~Treutner, J.-P. Gelas, L.~Lefevre, and A.-C. Orgerie.
\newblock Energy consumption side-channel attack at virtual machines in a
  cloud.
\newblock In {\em Dependable, Autonomic and Secure Computing (DASC), 2011 IEEE
  Ninth International Conference on}, pages 605--612. IEEE, 2011.

\bibitem{jana:2012}
S.~Jana and V.~Shmatikov.
\newblock Memento: Learning secrets from process footprints.
\newblock In {\em 2012 IEEE Symposium on Security and Privacy (S\&P '12)},
  pages 143--157, May 2012.

\bibitem{Killourhy:2009}
K.~Killourhy and R.~Maxion.
\newblock Comparing anomaly-detection algorithms for keystroke dynamics.
\newblock In {\em IEEE/IFIP International Conference on Dependable Systems
  Networks, 2009(DSN '09)}, pages 125--134, June 2009.

\bibitem{kim2012stealthmem}
T.~Kim, M.~Peinado, and G.~Mainar-Ruiz.
\newblock Stealthmem: System-level protection against cache-based side channel
  attacks in the cloud.
\newblock In {\em USENIX Security symposium}, pages 189--204, 2012.

\bibitem{kocher:introduction}
P.~Kocher, J.~Jaffe, and B.~Jun.
\newblock Introduction to differential power analysis and related attacks.
\newblock {\em
  \url{http://www.cryptography.com/resources/whitepapers/DPATechInfo.pdf}},
  1998.

\bibitem{kocher1999differential}
P.~Kocher, J.~Jaffe, and B.~Jun.
\newblock Differential power analysis.
\newblock In {\em Advances in Cryptology (CRYPTO '99)}, pages 388--397.
  Springer, 1999.

\bibitem{kocher:timing}
P.~C. Kocher.
\newblock Timing attacks on implementations of {Diffie-Hellman, RSA, DSS}, and
  other systems.
\newblock In {\em Advances in Cryptology (CRYPTO '96)}, pages 104--113.
  Springer, 1996.

\bibitem{Kopf:2010}
B.~K\"{o}pf and G.~Smith.
\newblock Vulnerability bounds and leakage resilience of blinded cryptography
  under timing attacks.
\newblock In {\em Proceedings of the 2010 23rd IEEE Computer Security
  Foundations Symposium}, CSF '10, pages 44--56. IEEE Computer Society, 2010.

\bibitem{li2014making}
D.~Li, A.~H. Tran, and W.~G. Halfond.
\newblock Making web applications more energy efficient for {OLED} smartphones.
\newblock In {\em Proceedings of the 36th International Conference on Software
  Engineering}, pages 527--538. ACM, 2014.

\bibitem{Li:2014:MWA}
D.~Li, A.~H. Tran, and W.~G.~J. Halfond.
\newblock Making web applications more energy efficient for {OLED} smartphones.
\newblock In {\em ICSE 2014}, pages 527--538, 2014.

\bibitem{lin2014screenmilker}
C.-C. Lin, H.~Li, X.~Zhou, and X.~Wang.
\newblock Screenmilker: How to milk your {Android} screen for secrets.
\newblock In {\em Proceedings of The 21th Annual Network and Distributed System
  Security Symposium ({NDSS})}, 2014.

\bibitem{mangard2008power}
S.~Mangard, E.~Oswald, and T.~Popp.
\newblock {\em Power analysis attacks: Revealing the secrets of smart cards},
  volume~31.
\newblock Springer Science \& Business Media, 2008.

\bibitem{messerges1999power}
T.~S. Messerges, E.~A. Dabbish, and R.~H. Sloan.
\newblock Power analysis attacks of modular exponentiation in smartcards.
\newblock In {\em Cryptographic Hardware and Embedded Systems}, pages 144--157.
  Springer, 1999.

\bibitem{messerges2002examining}
T.~S. Messerges, E.~A. Dabbish, and R.~H. Sloan.
\newblock Examining smart-card security under the threat of power analysis
  attacks.
\newblock {\em IEEE Transactions on Computers}, 51(5):541--552, 2002.

\bibitem{michalevsky2014gyrophone}
Y.~Michalevsky, D.~Boneh, and G.~Nakibly.
\newblock Gyrophone: Recognizing speech from gyroscope signals.
\newblock In {\em Proceedinga of 23rd USENIX Security Symposium, USENIX
  Association}, 2014.

\bibitem{MichalevskyNSB15}
Y.~Michalevsky, G.~Nakibly, A.~Schulman, and D.~Boneh.
\newblock {PowerSpy}: Location tracking using mobile device power analysis.
\newblock In {\em 24th USENIX Security Symposium (USENIX Security 15)},
  Washington, D.C., Aug. 2015. USENIX Association.

\bibitem{Miluzzo:2012}
E.~Miluzzo, A.~Varshavsky, S.~Balakrishnan, and R.~R. Choudhury.
\newblock Tapprints: Your finger taps have fingerprints.
\newblock In {\em Proceedings of the 10th International Conference on Mobile
  Systems, Applications, and Services}, MobiSys '12, pages 323--336. ACM, 2012.

\bibitem{moller2001securing}
B.~M{\"o}ller.
\newblock Securing elliptic curve point multiplication against side-channel
  attacks.
\newblock In {\em Information Security}, pages 324--334. Springer, 2001.

\bibitem{muller:dynamic}
M.~M{\"u}ller.
\newblock Dynamic time warping.
\newblock {\em Information retrieval for music and motion}, pages 69--84, 2007.

\bibitem{Qian:2012}
Z.~Qian, Z.~M. Mao, and Y.~Xie.
\newblock Collaborative {TCP} sequence number inference attack: How to crack
  sequence number under a second.
\newblock In {\em Proceedings of the 2012 ACM Conference on Computer and
  Communications Security}, CCS '12, pages 593--604. ACM, 2012.

\bibitem{ratanpal2004chip}
G.~B. Ratanpal, R.~D. Williams, and T.~N. Blalock.
\newblock An on-chip signal suppression countermeasure to power analysis
  attacks.
\newblock {\em Dependable and Secure Computing, IEEE Transactions on},
  1(3):179--189, 2004.

\bibitem{Saponas:2007}
T.~S. Saponas, J.~Lester, C.~Hartung, S.~Agarwal, and T.~Kohno.
\newblock Devices that tell on you: Privacy trends in consumer ubiquitous
  computing.
\newblock In {\em Proceedings of 16th USENIX Security Symposium on USENIX
  Security Symposium}, pages 5:1--5:16. USENIX Association, 2007.

\bibitem{schlegel:2011}
R.~Schlegel, K.~Zhang, X.-y. Zhou, M.~Intwala, A.~Kapadia, and X.~Wang.
\newblock Soundcomber: A stealthy and context-aware sound trojan for
  smartphones.
\newblock In {\em Proceedings of The 18th Annual Network and Distributed System
  Security Symposium ({NDSS})}, volume~11, pages 17--33, 2011.

\bibitem{Song:2001}
D.~X. Song, D.~Wagner, and X.~Tian.
\newblock Timing analysis of keystrokes and timing attacks on {SSH}.
\newblock In {\em Proceedings of the 10th Conference on USENIX Security
  Symposium - Volume 10}. USENIX Association, 2001.

\bibitem{placeraider-ndss13}
R.~Templeman, Z.~Rahman, D.~Crandall, and A.~Kapadia.
\newblock {PlaceRaider}: Virtual theft in physical spaces with smartphones.
\newblock In {\em Proceedings of The 20th Annual Network and Distributed System
  Security Symposium ({NDSS})}, Feb. 2013.

\bibitem{Vuagnoux:2009}
M.~Vuagnoux and S.~Pasini.
\newblock Compromising electromagnetic emanations of wired and wireless
  keyboards.
\newblock In {\em Proceedings of the 18th Conference on USENIX Security
  Symposium}, SSYM'09, pages 1--16. USENIX Association, 2009.

\bibitem{wang2007new}
Z.~Wang and R.~B. Lee.
\newblock New cache designs for thwarting software cache-based side channel
  attacks.
\newblock In {\em ACM SIGARCH Computer Architecture News}, volume~35, pages
  494--505. ACM, 2007.

\bibitem{wray1991analysis}
J.~C. Wray.
\newblock An analysis of covert timing channels.
\newblock In {\em Research in Security and Privacy, 1991. Proceedings., 1991
  IEEE Computer Society Symposium on}, pages 2--7. IEEE, 1991.

\bibitem{Wright:2008}
C.~V. Wright, L.~Ballard, S.~E. Coull, F.~Monrose, and G.~M. Masson.
\newblock Spot me if you can: Uncovering spoken phrases in encrypted {VoIP}
  conversations.
\newblock In {\em Proceedings of the 2008 IEEE Symposium on Security and
  Privacy}, pages 35--49. IEEE Computer Society, 2008.

\bibitem{wright:spycatcher}
P.~Wright and P.~Greengrass.
\newblock {\em {Spycatcher}: The candid autobiography of a senior intelligence
  officer}.
\newblock Dell Publishing Company, 1987.

\bibitem{Xu:2012}
Z.~Xu, K.~Bai, and S.~Zhu.
\newblock {TapLogger}: Inferring user inputs on smartphone touchscreens using
  on-board motion sensors.
\newblock In {\em Proceedings of the Fifth ACM Conference on Security and
  Privacy in Wireless and Mobile Networks}, WISEC '12, pages 113--124. ACM,
  2012.

\bibitem{yetim2012eprof}
Y.~Yetim, S.~Malik, and M.~Martonosi.
\newblock {EPROF: An} energy/performance/reliability optimization framework for
  streaming applications.
\newblock In {\em Design Automation Conference (ASP-DAC), 2012 17th Asia and
  South Pacific}, pages 769--774. IEEE, 2012.

\bibitem{yoon2012appscope}
C.~Yoon, D.~Kim, W.~Jung, C.~Kang, and H.~Cha.
\newblock {AppScope}: Application energy metering framework for {Android}
  smartphone using kernel activity monitoring.
\newblock In {\em USENIX Annual Technical Conference}, pages 387--400, 2012.

\bibitem{zhang2011predictive}
D.~Zhang, A.~Askarov, and A.~C. Myers.
\newblock Predictive mitigation of timing channels in interactive systems.
\newblock In {\em Proceedings of the 18th ACM conference on Computer and
  communications security}, pages 563--574. ACM, 2011.

\bibitem{Zhang:2009}
K.~Zhang and X.~Wang.
\newblock Peeping tom in the neighborhood: Keystroke eavesdropping on
  multi-user systems.
\newblock In {\em Proceedings of the 18th Conference on USENIX Security
  Symposium}, SSYM'09, pages 17--32. USENIX Association, 2009.

\bibitem{Zhang:2012}
Y.~Zhang, A.~Juels, M.~K. Reiter, and T.~Ristenpart.
\newblock {Cross-VM} side channels and their use to extract private keys.
\newblock In {\em Proceedings of the 2012 ACM Conference on Computer and
  Communications Security}, CCS '12, pages 305--316. ACM, 2012.

\bibitem{zhou:2013}
X.~Zhou, S.~Demetriou, D.~He, M.~Naveed, X.~Pan, X.~Wang, C.~A. Gunter, and
  K.~Nahrstedt.
\newblock Identity, location, disease and more: Inferring your secrets from
  {Android} public resources.
\newblock In {\em Proceedings of the 2013 ACM SIGSAC Conference on Computer
  Communications Security}, CCS '13, pages 1017--1028. ACM, 2013.

\bibitem{zhou2012dissecting}
Y.~Zhou and X.~Jiang.
\newblock Dissecting android malware: Characterization and evolution.
\newblock In {\em IEEE S\&P}, pages 95--109. IEEE, 2012.

\bibitem{Zhuang:2009}
L.~Zhuang, F.~Zhou, and J.~D. Tygar.
\newblock Keyboard acoustic emanations revisited.
\newblock {\em ACM Transaction Information System Security}, 13(1):3:1--3:26,
  Nov. 2009.

\end{thebibliography}

\end{document}